\documentclass[final,superscriptaddress,english,twocolumn,amssymb, nobibnotes, aps, PRB,longbibliography,nofootinbib]{revtex4-1}

\usepackage{graphicx}
\usepackage{float}
\newcommand{\tr}{\mathrm{Tr}}
\usepackage{amsmath}
\usepackage{amssymb}
\usepackage{appendix}
\usepackage[dvipsnames]{xcolor}
\usepackage{tikz}
\usepackage[normalem]{ulem}
\usepackage[section]{placeins}
\usepackage{ifthen}
\definecolor{darkblue}{rgb}{0,0,.65}
\definecolor{darkgreen}{rgb}{0.28,0.41,0.19}
\usepackage[bottom]{footmisc}

\usepackage[%
  pdfauthor={Kevin Hemery},%
  pdfstartview=FitH,%
  breaklinks=true,%
  bookmarks=true,%
  colorlinks=true,%
  anchorcolor=black,%
  citecolor=blue,
  filecolor=black,%
  menucolor=black,%
  urlcolor=darkblue,%
  linkcolor=blue,%
 ]{hyperref}
\usepackage[all]{hypcap} 
\newcommand{\bra}[1]{\langle\,#1\,|}
\newcommand{\ket}[1]{|#1\rangle}

\renewcommand{\vec}[1]{\boldsymbol{#1}}

\newcommand{\norm}[1]{\left \| #1 \right \|}

\definecolor{nicegreen}{RGB}{0,100,0}
 
\begin{document}

\title{Matrix product states approaches to operator spreading in ergodic quantum systems}%

\author{K\'evin H\'emery}
\affiliation{Department of Physics T42, Technische Universit\"at M\"unchen, James-Franck-Stra\ss e 1, 85748 Garching, Germany}
\author{Frank Pollmann}
\affiliation{Department of Physics T42, Technische Universit\"at M\"unchen, James-Franck-Stra\ss e 1, 85748 Garching, Germany}
\affiliation{Munich Center for Quantum Science and Technology (MCQST), Schellingstr. 4, D-80799 München}
\author{David J. Luitz}
\affiliation{Max Planck Institute for the Physics of Complex Systems, Noethnitzer Str. 38, Dresden, Germany}
\email{dluitz@pks.mpg.de}
\date{December 14, 2018}%

\begin{abstract}
    We review different matrix product states (MPS) approaches to study the spreading of operators in generic nonintegrable quantum systems. As a common ground to all methods, we quantify this spreading by means of the Frobenius norm of the commutator of a spreading operator with a local operator, which is usually referred to as the out-of-time-order correlation (OTOC) function. We compare two approaches based on matrix-product states in the Schr\"odinger picture: the time dependent block decimation (TEBD) and the time dependent variational principle (TDVP), as well as TEBD based on matrix-product operators directly in the Heisenberg picture. The results of all methods are compared to numerically exact results using Krylov space exact time evolution. We find that for the Schr\"odinger picture the TDVP algorithm performs better than the TEBD algorithm.  
    Moreover the tails of the OTOC are accurately obtained both by TDVP MPS and TEBD MPO. They are in very good agreement with exact results at short times, and appear to be converged in bond dimension even at longer times. However the growth and saturation regimes are not well captured by neither of the methods.
\end{abstract}

\maketitle

\section{Introduction}
The question of quantum thermalization in closed systems receives currently a considerable amount of attention. This interest is partly due to the importance of this problem for the understanding of the foundations of statistical physics \cite{ETH_deutsch,Srednicki, Rigol, rigol_review, borgonovi_quantum_2016} and to the experimental progress leading to increasingly well isolated experimental realizations of quantum many-body systems in ultracold atomic gases in optical lattices \cite{zwerger_bloch_review}.
In general, the unitary dynamics of isolated quantum systems precludes reaching a maximally mixed state if the system is initialized in a pure state. Nevertheless, generic isolated quantum many-body systems typically are found to reach a thermal state, i.e. local observables assume values consistent with the micro-canonical ensemble at long times\cite{Kaufman794}.
Considering only a small subsystem of the total system, the usual notion of thermodynamic equilibrium is recovered, as the reduced density matrix becomes equal to the corresponding thermodynamic density matrix \cite{garrison_does_2018,luitz_long_2016}, as the rest of the system serves as a heat bath.
This behavior is rooted in the local structure of eigenstates of the many body Hamiltonian 
which manifests itself in the eigenstate thermalization hypothesis (ETH) \cite{ETH_deutsch, Srednicki, Rigol,rigol_review, borgonovi_quantum_2016}, motivated by random matrix theory considerations. It provides a precise prediction for the matrix form of local operators in the eigenbasis of the Hamiltonian, and implies thermalization\cite{Rigol,luitz_anomalous_2016, roy_anomalous_2018}.
The mechanism for this thermalization process is the loss of local quantum information over time, which implies that the full wave function of the initial state cannot be reconstructed from local measurements at long times \cite{Huse_and_Nandkishore}. 
In consequence of this loss of local information, the system becomes increasingly entangled, until the state of a subsystem reaches a maximally mixed state consistent with global constraints \cite{non_integrable_kim_huse,luitz_extended_2016,Kaufman794}. 
%
A direct local probe of the loss of local quantum  information can be constructed by studying the spreading of initially local Heisenberg operators $\hat{O}_i(t)$ which become increasingly nonlocal over the course of time. The locality can be quantified by probing the real space support of $\hat{O}_i(t)$ using the norm of the commutator with another local operator $\hat{V}_j$. This quantity is now best known as the out-of-time-order correlator (OTOC) which was introduced to study quantum chaos \cite{Larkin_Ovchinnikov, a_bound_to_chaos}, and to bound the spreading of information in systems with short ranged interactions \cite{Lieb1972}.

Certain universal properties of the OTOC can be well understood in random unitary circuits \cite{Curt_tibor,tibor_Curt_Charges, khemani,Nahum18} where it is governed by hydrodynamic equations of motion. In these systems, a light cone structure was identified with a broadening front arising from the diffusive nature of the hydrodynamic equations. This diffusive behaviour has been found in numerically exact calculations in a noisy spin system \cite{Michael_disorder}. However, no exponential regime with a fixed Lyapunov exponent was found so far in such systems, nor in Hamiltonian systems with a small local Hilbert space and continuous time\cite{luitz_information_2017}. 

While the OTOC is a powerful and universal theoretical tool, it is very difficult to calculate in practice for generic quantum many-body systems, due to its operator nature (see paragraph \ref{subsec:intro_otoc}). In the last two years several numerical methods for calculating the OTOC emerged: exact operator evolution in the Heisenberg picture \cite{chen_out--time-order_2016}, matrix product operators (MPO) evolution in the Heisenberg picture \cite{bohrdt_scrambling_2016,Xu_Swingle} and an exact wave function technique in the Schr\"odinger picture \cite{luitz_information_2017}. In this article, we will carefully compare these techniques and add two more MPS methods based on a stochastic sampling of the OTOC in the Schr\"odinger picture using both time evolving block decimation (TEBD) \cite{TEBD2003} and the time dependent variational principle (TDVP) using matrix product states (MPS) \cite{TDVP_old,TDVP_new}, which is currently discussed as a candidate method to extract late time hydrodynamic properties of quantum systems \cite{Leviatan_thermalization_mps,yevgeny}. While exact Schr\"odinger evolution using quantum typicality is currently the best choice to obtain the exact OTOC for Hilbert space dimensions of up to $10^9$  even at late times \cite{luitz_information_2017}, MPS techniques have been recently presented as complementary approaches. In particular, MPO time evolution using TEBD can be used to extract the tails of the OTOC at very long distances (see paragraph \ref{subsec:intro_otoc}). Here, we investigate, how MPS based time evolution techniques such as TEBD and TDVP in the Schr\"odinger picture compare to the method of MPO evolution.

This article is structured as follows: in section \ref{section:tMPS} we review the different MPS methods that we used to simulate quantum dynamics. In section \ref{sec:chaos}, we present different ways of quantifying the spreading of operators in non-integrable systems, and the numerical approaches we choose to simulate them.
Next, in section \ref{results} we compare the results obtained by the different methods, both for small and larger systems, and assess to which extent our results can be trusted despite the low amount of entanglement included in our MPS approximations.

\section{Matrix product states methods for the simulation of quantum time evolution}
\label{section:tMPS}
While the MPS formalism originally arose in the context of ground state physics, it has also been very successful in the description of quantum dynamics of one dimensional quantum systems \cite{SCHOLLWOCK}. Every quantum many body state can be brought into a MPS form, that is:
\begin{equation}
\ket{\psi}=\sum_{s_1,...,s_N} A^{[1] s_1}A^{[2] s_2}...A^{[N]s_N}\ket{s_1,s_2,...,s_N}
\label{eq:MPS}
\end{equation}
where $A^{[i]s_i}\in \mathbb{C}^{\chi\times\chi}$ is the matrix corresponding to site $i$ and to the local state $\ket{s_i}$. Note that $A^{[1]s_1}$ ($A^{[N]s_N}$) are row (column) vectors of size $\chi$ to make the wavefunction coefficients scalar. The required bond dimension $\chi$ depends on the amount of bipartite entanglement contained in $\ket{\psi}$. In this section, we review different MPS based method to perform quantum time evolution.
\subsection{Time evolving block decimation}

The TEBD algorithm \cite{TEBD2003,SCHOLLWOCK} is a powerful and simple tool to simulate the short time dynamics of one-dimensional systems with short range interactions. It is applicable to both the Schr\"odinger and Heisenberg picture. 

\begin{figure}
	\centering
	\includegraphics{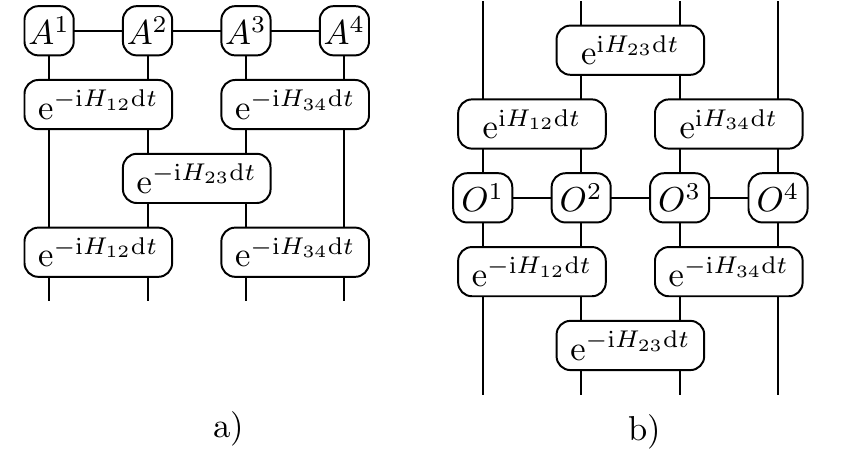}
\caption{ Schematic representation of the TEBD algorithm in a schematic notation: a) Schr\"odinger representation, where $A^i$ is the tensor at site $i$ of the MPS corresponding to the state that we wish to evolve, $H_{ij}$ denotes the gates of the Hamiltonian between sites $i$ and $j$ and $\mathrm{d}t$ is the time step; b) Heisenberg representation, where $O^i$ is the tensor at site $i$ of the MPO corresponding to the operator that we wish to evolve.}
	\label{fig:full_tebd}
\end{figure}

The key idea of the algorithm is to employ a Trotter decomposition of the time evolution operator in such a way that only local time evolution operators occur. For example a nearest-neighbour Hamiltonian is written as a sum over terms which involve only two neighboring sites: $\hat{H}=\sum_{i}h_{i,i+1}$ where $h_{i,i+1}$ is the part of $\hat{H}$ containing the interaction between sites $i$ and $i+1$. Since only nearest neighbor terms do not commute, we decompose the Hamiltonian into a part acting on the even bonds and a part acting on the odd bonds as $\hat{H}=H_{\mathrm{even}}+H_{\mathrm{odd}}$ where $H_{\mathrm{even}}=\prod_{j}h_{2j,2j+1}$ and $H_{odd}=\prod_{j}h_{2j+1,2j+2}$. In this way all the terms contained in $H_{\mathrm{even}}$ (resp. $H_{\mathrm{odd}}$) commute with each other. We can now apply a Trotter decomposition at first order for simplicity, although the implementation of any order is possible within this scheme. We obtain for the time evolution operator:
\begin{equation}
\begin{split}
U(\mathrm{d}t)&=\mathrm{e}^{-\mathrm{i}(H_{\mathrm{odd}}+H_{\mathrm{even}}) \mathrm{d} t} \\&\approx \mathrm{e}^{-\mathrm{i} H_{\mathrm{odd}} \mathrm{d}t} \mathrm{e}^{-\mathrm{i} H_{\mathrm{even}} \mathrm{d}t}+O(\mathrm{d} t^2)\\ 
&=\prod_{j}\mathrm{e}^{-\mathrm{i} h_{2j,2j+1}\mathrm{d}t}\prod_{j}\mathrm{e}^{-\mathrm{i} h_{2j+1,2j+2}\mathrm{d}t}+O(\mathrm{d}t^2)
\label{eq:gates}
\end{split}
\end{equation}
Each term of the products in the third line of equation \eqref{eq:gates} can be written as a unitary gate linking two adjacent sites.
In order to evolve an MPS with a time step $\mathrm{d}t$ we apply a layer of gates as shown in Fig. \ref{fig:full_tebd}. It is then possible to re-express the time evolved state as an MPS of higher bond-dimension. 

After each application of a unitary gate, the MPS is optimized using the Schmidt decomposition.
This is done efficiently by writing the state in a Schmidt basis, and truncating the smallest Schmidt values, whose contributions to the wave-function are least important. This approximation is only valid in a regime where the state is lowly entangled, limiting the use of the method to short times for ergodic systems such as the ones we are studying in this work.
Moreover the truncation process renders the time-evolution non unitary and does not preserve the norm of the state, the energy and other conserved quantities. 

The TEBD algorithm is straightforwardly extended to the MPO-time evolution as illustrated on Fig. \ref{fig:full_tebd}b, by noting that in contrast with the MPS case the (adjoint) unitary time evolution must be applied on the lower and upper legs of the MPO.

All the results obtained with the TEBD algorithm presented in this article were performed with a second order Trotter decomposition scheme. 

\subsection{The time dependent variational principle using matrix-product states}

The time dependent variational principle (TDVP) was first introduced by Dirac \cite{dirac_1930} for a general variational manifold. The general idea is to project the Schr\"odinger equation on the variational manifold of interest in such a way that the wave function after an infinitesimal time step does not leave the manifold, yielding an approximate tractable time evolution. 

It has been recently formulated using MPS with a fixed bond dimension as the variational manifold \cite{TDVP_new,TDVP_old} and 
is based on the concept of the tangent space of the MPS manifold \cite{Post-mps}.
The algorithm is very similar to the density matrix renormalization group (DMRG) \cite{DMRG1,equivalence_DMRG} and offers several advantages with respect to the TEBD algorithm since it does not rely on truncation to keep the wave function in the manifold $\mathcal{M_{MPS,\chi}}$ of MPS with a given dimension $\chi$. Moreover it is suitable to simulate Hamiltonians with long range interactions.
To derive the TDVP algorithm, we again start from an MPS wave function Eq. \eqref{eq:MPS} as a variational ansatz. However, we make every tensor explicitly time dependent, i.e. $A^{[i]s_i} \rightarrow A^{[i]s_i}(t)$.
 
After inserting this ansatz into the Schr\"odinger equation, we find:
\begin{equation}
   \sum_i \dot{A}^{[i]s_i}(t) \partial_{A^{[i]s_i}} \ket{ \psi(\{A^{[k]s_k}(t)\}}=-i P_T H \ket{\psi(\{A^{[k]s_k}(t)\})}.
\end{equation}
where $P_T$ is the projector on the tangent plane of $\ket{\psi(\{A^{[k]s_k}(t)\})}$. This projection  is necessary to ensure that we obtain  closed equations of motion in the tangent space so that the time evolved MPS is confined to $\mathcal{M_{MPS,\chi}}$. The new equations of motion obtained this way can be elegantly integrated by means of a splitting method. For more details, we refer to ref. \cite{TDVP_new}. 

Unlike the TEBD algorithm, the TDVP algorithm conserves energy even when the exact time evolution cannot be captured by the MPS. Moreover all the conserved quantities, which do not cause an increase of bond dimension once applied to a MPS, are respected. In other words, a quantity $\hat{O}$ commuting with the Hamiltonian is conserved by TDVP if for all states $\ket{\phi}$ expressible as a MPS of bond dimension $\chi$ we can still express $\hat{O} \ket{\phi}$ as a MPS of bond dimension $\chi$.
In the case where all conserved quantities leave the manifold  $\mathcal{M_{MPS,\chi}}$ invariant, TDVP appears well suited for the simulation of thermalization, since it is believed that at long times the dynamics of the system is driven by hydrodynamical equations of motion governed by conserved quantities \cite{lecture_hydrodynamics,quantum_hydrodynamics,Lucas_hydrodynamics}. Following this line of thought, TDVP has recently been applied successfully in the context of thermalization \cite{Leviatan_thermalization_mps}.
However the accuracy of results at long times is currently under debate \cite{yevgeny}. It has also been used advantageously in disordered systems \cite{Elmer_Doggen}. In this work, we focus on short to intermediate times and will not consider the question of the relevance of TDVP in the context of hydrodymamics. 
However the TDVP algorithm provides a unitary time evolution. This feature will be of crucial importance when calculating OTOCs, as explained in section \ref{subsec:intro_otoc}. 
\section{Measures of operators spreading in closed quantum systems: the out-of-time-order correlator}
\label{sec:chaos}
\subsection{Definition}
\label{subsec:intro_otoc}
In ergodic isolated quantum systems, local operators in the Heisenberg picture typically spread over the course of time in a sense that their supports in real space grows. Here we study this spreading in a non-integrable spin-$\frac{1}{2}$ chain of length $L$. In this case, the growth of the support means that the expansion of an operator $\hat V_i(t)$ in the operator basis of strings of local Pauli operators $\hat \sigma_i$, $i\in \{0,1,2,3\}$ ($\hat \sigma_0 = \hat 1$):
\begin{equation}
    \hat V_i(t) = \sum_{\alpha_1\dots\alpha_L\in\{0,1,2,3\}} v_{\alpha_1,\dots,\alpha_L} \hat \sigma_{\alpha_1} \otimes \hat \sigma_{\alpha_2}\otimes \dots \otimes \hat \sigma_{\alpha_L}
    \label{eq:op_expansion_pauli}
\end{equation}
acquires increasingly long strings of non-identity Pauli operators \cite{Curt_tibor,khemani,Nahum18}. The growing complexity stems from  the increasing nonlocality of the time evolution operator $\hat U(t)$ and is also related to the growth of the \textit{operator entanglement entropy} \cite{pizorn_operator_2009,zhou_operator_2017,dubail_entanglement_2017}. In order to define the latter quantity, it is useful to consider the operator as a state living in a larger Hilbert space. More precisely, if we decompose an operator $\hat{O}$ in terms of the elements of an orthonormal basis $ \left \{ \ket{\phi_i} \right \} $ as $\hat{O}=\sum_{ij} O_{ij} \ket{\phi_i}\bra{\phi_j}$, we can then associate to this operator the state $\ket{\phi_{\hat{O}}}  =\sum_{ij} O_{ij}\ket{\phi_i}\otimes\ket{\phi_j}$. The bipartite entanglement entropy of $\ket{\phi_{\hat{O}}}$ is called the operator entanglement entropy (assuming proper normalization). Note that the calculation of the entanglement of MPOs is identical to the MPS case: one has to find the singular values of the bond $i$ of interest and as usual $S=-\sum_{j} s_j^2 \ln(s_j^2)$ where $s_j$ are the singular values and $S$ the entanglement. 

The expansion in terms of Pauli string operators in Eq. \eqref{eq:op_expansion_pauli} is a useful measure of operator spreading but is computationally impractical due to the arising of an exponentially large number of terms in the length of the chain. However, we note that in order to study the loss of locality of quantum information, the most interesting information is contained in how the length of Pauli strings grows over time. Therefore, we instead consider \textit{commutators} of the operator $\hat V_i(t)$ with nontrivial ($\alpha\in\{x,y,z\}$) local Pauli operators (i.e. $\hat \sigma_\alpha^i = \hat 1 \otimes \dots \otimes \hat 1 \otimes \hat \sigma_\alpha \otimes\hat 1 \otimes \dots \otimes\hat 1$):

\begin{equation}
    \left[ \hat V_i(t), \hat \sigma_\alpha^j\right]  =  \sum_{\alpha_1\dots\alpha_L} v_{\vec \alpha} \left[ \hat \sigma_{\alpha_1} \otimes \hat \sigma_{\alpha_2}\otimes \dots \otimes \hat \sigma_{\alpha_L},  \hat \sigma_\alpha^j \right].
\end{equation}

This commutator is zero when the local Pauli operators $\hat{\sigma}_{\alpha_j}$ on site j are identities for all strings in the expansion $\hat V_i(t)$. Generically, when the operator support of $\hat V_i(t)$ reaches site $j$, its expansion in terms of Pauli strings will contain terms not commuting with $\hat{\sigma}_\alpha^j$. This commutator is therefore quantifying the operator spreading.
However it is also an operator and is therefore usually reduced to its norm $\norm{\left[ \hat V_i(t), \hat \sigma_\alpha^j\right]}$. For computational simplicity, a standard choice for the norm is the normalized Frobenius norm given by $\norm{\hat A}_F^2 = \frac{1}{\mathcal N} \tr \hat A^\dagger \hat A$, leading to the definition
\begin{equation}
    C_{ij}(t)=\frac{1}{2 \cdot Z} \left\| [\hat W_j,\hat V_i(t)] \right\|_F^2,
\label{eq:otocdef}
\end{equation}
where $Z$ is the dimension of the Hilbert space. 
We have chosen the normalization in order to ensure that $C_{ij}(t) \to 1$ at long times in the case where $\hat W_j$ and $\hat V_i$ are hermitian operators, which square to identity, such as Pauli operators.

This quantity was originally proposed by Larkin and Ovchinniokv \cite{Larkin_Ovchinnikov} in the context of quantum chaos. They showed that in chaotic systems with a semiclassical limit this norm of the commutator is connected to a Lyapunov exponent of the system and therefore effectively quantifies its chaoticity. Using this quantity they discussed a quantum analogue of classical chaos since in the semi-classical limit it quantifies the sensibility of classical trajectories to their initial conditions for the choice $\hat{W}_j=\hat{p}$ and $\hat{V}_i=\hat{x}$. This can be understood more intuitively by observing that the OTOC measures the effect of an initial perturbation on the value at later times of an operator located at some distance \cite{a_bound_to_chaos}.
On the other hand, recent numerical studies of quantum systems with a small local Hilbert space and for $31$ sites
  showed that there is no regime of exponential growth \cite{luitz_information_2017}, a discrepancy to the semiclassical case \cite{rammensee_many-body_2018}, which has yet to be fully understood.

The link between the OTOC and locality of the Hamiltonian was made a few years later by Lieb and Robinson\cite{Lieb1972}. They realized that information in systems with short range interactions can only spread within a light-cone with only exponentially suppressed leaking. This is most effectively quantified by considering the spreading of initially local operators $\hat V_i(t)$ in the Heisenberg picture.
More precisely:
\begin{equation}
    \lim_{t\to\infty,|i-j|>v t} C_{ij} \exp{\left [ \mu(v) t \right ]} = 0,
    \label{eq:lieb_robinson}
\end{equation}
for velocities $v>v_{LR}$, where $v_{LR}$ is called the Lieb-Robinson velocity. The function $\mu(v)$ is now referred to as velocity dependent Lyapunov exponent \cite{khemani_velocity-dependent_2018}.

The OTOC has been the subject of a renewed interest in the past few years due the establishment of a duality between some strongly correlated systems and black-holes and the proposal of exactly solvable models to illustrate it \cite{Kitaev}. Moreover the spreading of operators is directly connected to the scrambling of local quantum information, since in chaotic systems at long times, initially local operators lose their locality and become completely scrambled \cite{a_bound_to_chaos}. 

\subsection{Numerical considerations}
\label{sec:numerical consideration}
If we restrict ourselves to hermitean unitary operators, which square to identity, such as Pauli operators, the OTOC $C_{ij}$ can be expressed as:
\begin{equation}
C_{ij} =  1 - \frac{1}{2 \cdot Z} \tr\left(\hat{V}_i(t) \hat{W}_j \hat{V}_i(t) \hat{W}_j \right),
\label{eq:otocdef}
\end{equation}

where $Z=\mathrm{dim}(\mathcal H)$ is the dimension of the Hilbert space (see paragraph \ref{subsec:intro_otoc}).
The nontrivial part of the calculation of this quantity consists of determining the correlation function
\begin{equation}
\frac{1}{Z} \tr\left( \hat{V}_i(t) \hat{W}_j \hat{V}_i(t) \hat{W}_j \right),
\label{eq:otoc}
\end{equation}
which is exponentially expensive: for a spin-$\frac 1 2 $ system of size $L$, $\hat{V}_i(t)$ is represented by a matrix in $\mathbb{C}^{2^{L}\times 2^{L}}$.
Direct exact time evolution of the operator will therefore be very limited in system size \cite{chen_out--time-order_2016}.

Alternatively the trace can be stochastically evaluated with typical, randomly chosen wave functions due to quantum typicality. Although the time evolution will still be exponentially expensive, larger system sizes can be achieved since a state has only $2^L$ components. This has been achieved using exact Krylov space time evolution \cite{luitz_information_2017,luitz_emergent_2018,Krylov1,Krylov2,exponentialmatrix}. Here, instead of operators, only wave functions are evolved in time by moving to the Schr\"odinger picture at the price of performing the time evolution forward and backwards in time, yielding an overall scaling of the method proportional to $t_\text{max}^2$. This is so far the most powerful numerically exact method to simulate the ergodic dynamics of small to intermediate system sizes up to arbitrary times and used here as a benchmark. For details of the method, see Refs. \onlinecite{luitz_information_2017, luitz_emergent_2018}.

\begin{figure}
 	\centering
 	\includegraphics{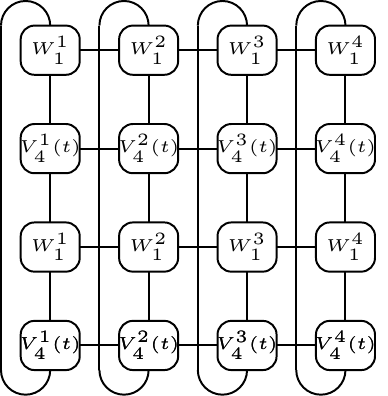}
 	\caption{Example of trace calculation with MPOs following equation \eqref{eq:otocdef}.}
 	\label{fig:trace}
 \end{figure}

Another approach is to use Heisenberg propagation of a matrix product operator (MPO) representation \cite{bohrdt_scrambling_2016,Xu_Swingle} of $\hat{V}_j(t)$ (see Fig. \ref{fig:full_tebd}).
 To calculate $C_{i,j}$, one can use equation \eqref{eq:otocdef}. We first evolve $\hat{V}_j$ using the setup of Fig.\ref{fig:full_tebd}b) with $O^j=\hat{V}$ and all the other operators $O^k$, $k \neq j$ set to identity. This way we obtain an MPO which tensors we denote as $V_j^l(t)$, the index $l$ corresponding to the site of the tensor. We also write $W_i$ as a MPO (which means that we place the operator $\hat{W}$ on site $i$ and identities operators on every other site). The calculation of the trace is then performed according to Fig.\ref{fig:trace}.
 This approach can seem to be limited to short times due to the linear growth of the operator entanglement entropy \cite{pizorn_operator_2009} implying the necessity of an exponentially large bond dimension for an exact representation of the Heisenberg operator.

\begin{figure}
\includegraphics{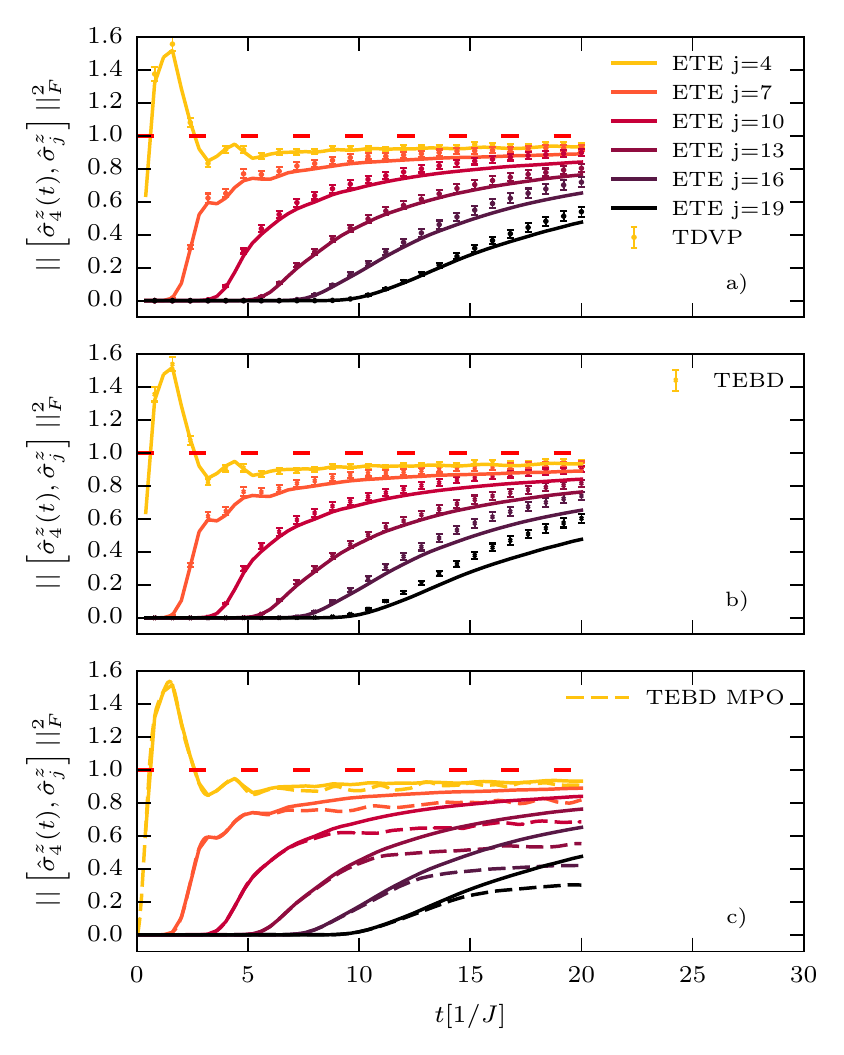}
\caption{Fixed space cuts of the OTOC $C_{4,j}=\frac{1}{2}\|\left[ \hat \sigma_4^z(t), \sigma_j^z\right]\|_F^2$ obtained using different methods compared to exact time evolution (ETE). a) TDVP with wave function time evolution, where the trace in Eq. \eqref{eq:otocdef} is sampled using 98 random product states b) TEBD with wave function time evolution where the trace is sampled using 98 random states (the errorbars represent the error coming from the stochastic sampling) c) TEBD with operator time evolution (exact trace). All MPS calculations were performed with a time step $\mathrm{d}t=0.01$, bond dimension $\chi=64$ and system size $L=21$. The red dashed line is the theoretical upper bound for the long time limit for completely scrambled operators.}
\label{fig:panel_cuts}
\end{figure}

 It has been argued recently in Ref. \cite{Xu_Swingle} that this method is able to capture the early growth of the OTOC even with low bond dimension based on the following observation: first the spreading of quantum information is bounded by a light cone, implying that the operator entanglement of bipartitions with a cut outside of the light cone is small, thus leading to a small required bond dimension. Therefore in the Heisenberg picture, only tensors inside the region where the entanglement is high will be truncated within the TEBD scheme. Finally, it is assumed that the effect of the truncation propagates as a light cone, meaning that the sites with low entanglement should not be affected immediately by the effect of a truncation far away from them. In numerical simulations, the convergence of the results with bond-dimension presented in Ref. \cite{Xu_Swingle} seems to support this reasoning. Our benchmarks for small systems (Fig. \ref{fig:panel_cuts}), our comparison of the contour lines of the OTOC obtained using MPO evolution to other methods (Fig. \ref{fig:comparison_contours}), as well as our analysis of the convergence with bond dimension (Fig. \ref{fig:convergence chi MPO}) also provide further support that MPO time evolution does indeed accurately capture the tail of the OTOC. 
 We would like to note that this MPO technique has been applied in the past to calculate operator spreading in the one-dimensional Bose-Hubbard model in Ref. \cite{bohrdt_scrambling_2016}, where a discrepancy to the ballistic spreading at early times for small bond dimensions was pointed out, which was attributed to the truncation of the bond dimension.

Here, we propose a scheme based on the Schr\"odinger picture to MPS time evolution methods (see section \ref{section:tMPS}). The trace in Eq. \eqref{eq:otoc} is sampled stochastically over random product states $\ket{\sigma_1,\dots\sigma_L}$, which is reminiscent of  minimally entangled typical thermal states (METTS) at infinite temperature\cite{white_minimally_2009} ($\beta=0$). Additionally, we have the freedom to chose the basis such that the basis states are eigenstates of the operator $\hat{W}_j$, which we
take for convenience to be $\hat{W}_j=\hat{\sigma}_i^z$: 
\begin{equation}
\begin{split}
&\frac{1}{Z} \tr\left( \hat{V}_i(t) \hat{\sigma}_z \hat{V}_i(t) \hat{\sigma}_z \right) \approx \\
&\approx
\frac{1}{Z \cdot n_\text{states}} \sum_{\vec{\sigma}}^{n_\text{states}} \bra{\vec{\sigma}} \hat{V}_i(t)
\hat{\sigma}^z_j \hat{V}_i(t) \ket{\vec{\sigma}} \sigma_j. \\
\end{split}
\label{eq:metts_otoc}
\end{equation}

This way, we only have to propagate one wave function (i.e. $\ket{\vec{\sigma}}$) forward in time,
apply $\hat{V}_i$, and propagate back to $t=0$ for each initial state $\ket{\vec{\sigma}}$. The
average is performed over $n_\text{states}$ initial states, which are sampled uniformly from the
local $\sigma_z$ product state basis (subject to sector constraints if required). From now on, we we will restrict ourselves to the case $\hat{V}_i(t)=\hat{\sigma}_i^z(t)$.

In order to evaluate equation \eqref{eq:metts_otoc}, we use both the TEBD and the single site TDVP algorithms. In ref. \cite{TDVP_new}, a two-site implementation of the TDVP algorithm was proposed. However, in our case, this version of the algorithm would not be suitable for our purposes since it also relies on truncation to keep the bound dimension of the MPS fixed, hence yielding a non unitary time evolution and violating conservation of energy. However the single site algorithm does not allow to increase dynamically the bond dimension. In order to address this problem,  we initialize our MPS as follows. First, we fill up the MPS with zeros in such a way that the product state, initially of bond dimension one, acquires the desired bond dimension $\chi$. Second, we bring our inflated product state in canonical (isometric) form following the usual sweeping procedure \cite{SCHOLLWOCK}. This state is then a proper state and the single-site TDVP reproduces the correct time evolution.

\section{ Results}
\label{results}

\begin{figure}
\includegraphics{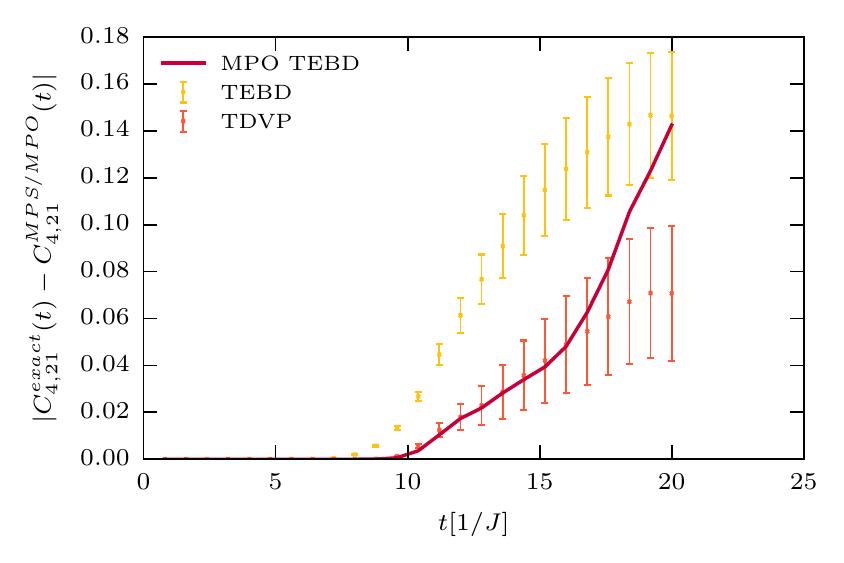}
\caption{Deviation from the exact result using the three methods for the calculation of the OTOC for the longest distance from site $i=4$ to $j=21$ as a function of time. Based on the results shown in Fig. \ref{fig:panel_cuts} (Bond dimension $\chi=64$, time step $\mathrm{d}t=0.01$, system size $L=21$ sites, the errorbars represent the error coming from the stochastic sampling).}
\label{fig:errors_methods}
\end{figure}

We study a one dimensional quantum spin chain with short range interactions which has both integrable
and non-integrable points as a function of the field angle: the tilted field Ising model.
The Hamiltonian of the system is given by:
\begin{equation}
H=\sum_{i=1}^{L-1} J_z \hat{\sigma}^z_i \hat{\sigma}^z_{i+1} +\sum_{i=1}^L \left(h_x \hat{\sigma}^x_i+h_z
\hat{\sigma}^z_i \right)
\end{equation}
We consider this model at a strongly nonintegrable point, with no other conservation laws besides the global conservation of energy, which is exactly respected by our TDVP approach. Following Ref. \cite{non_integrable_kim_huse}, we use the following parameters throughout this article:
$J_{z}=1$, $h_x=(\sqrt{5}+1)/4=0.8090$,$h_z= (\sqrt{5}+5)/8=0.9045$.

\subsection{Comparison of the methods with exact results }

\label{sec:benchmarks}
In order to explore the domain of validity of the different methods (MPS TDVP, MPS TEBD and MPO TEBD), we compare the results for the OTOC with exact results obtained by Krylov space based exact time evolution (ETE) for chains of $L=21$ spins. 

In Fig. \ref{fig:panel_cuts}, we present a detailed comparison for a system of size $L=21$ of the OTOC $C_{4,j}=\frac{1}{2}\|\left[ \hat \sigma_4^z(t), \sigma_j^z\right]\|_F^2$ obtained from the four methods compared in this article.
All panels show the numerically exact result obtained from exact time evolution (ETE) as solid lines, panel a) shows the TDVP result for the OTOC obtained from stochastic sampling of the trace in Eq. \eqref{eq:otocdef} using 98 random product states, panel b) shows the same calculation but using TEBD time evolution instead. 
In panel c), we show TEBD MPO evolution results, using a direct evaluation of the trace. Here, all calculations where performed using a maximal bond dimension of $\chi=64$. 
It is clear that at short times all three methods reproduce the exact result since there is no significant truncation occurring. 
Interestingly, the TDVP results stay close to the exact result for longer times than the OTOC obtained by TEBD. 
Similarly to TDVP, the MPO evolution using TEBD captures very well the regime of low values of the OTOC. Nevertheless the growth and saturation regime is not correctly reproduced by any of the methods. With the Schr\"odinger approach the OTOC is systematically overestimated while it saturates to an unphysical value in the case of the Heisenberg approach. 

In order to make these statements more precise, we investigate the error of the different methods by considering the deviation from the exact result for $L=21$. From Fig. \ref{fig:panel_cuts}, we see that the discrepancy from the exact result is the largest at long times and long distances, independently of the choice of the method. Therefore we illustrate the errors resulting from each of the three methods at the longest spatial distance in our system from the origin at $i=4$ by the distance to the exact result $|C^\text{exact}_{4,21}(t) - C^\text{MPS/MPO}_{4,21}(t)|$ in Fig. \ref{fig:errors_methods}.
Here $C^\text{MPS/MPO}_{4,21}(t)$ stands for the OTOC calculated using either the MPS based methods with TEBD or TDVP time evolution or by the direct MPO based approach. We have checked that similar results are obtained for other distances.

These results are explained as follows. While TEBD in the Schr\"odinger picture suffers from the propagation of truncation errors since the wave function has to be propagated back to $t=0$ after a measurement, TDVP profits from the preservation of unitarity by the method, leading to a significantly smaller error compared to TEBD MPS.
As for the Heisenberg picture, the low value of operator entanglement at the front of the OTOC light cone allows the MPO TEBD approach to capture the low values of the OTOC as explained in paragraph \ref{sec:numerical consideration}.
The systematic underestimation of the OTOC saturation values is due to finite bond dimensions limiting the captured operator entanglement. We note that similar results have been obtained in disordered spin chains where it has been demonstrated that TDVP performs better than TEBD \cite{Hubig_review}.

\subsection{Large systems and range of validity of the approximation}
\label{sec:absence of broadening}

	\begin{figure}
\centering
\includegraphics{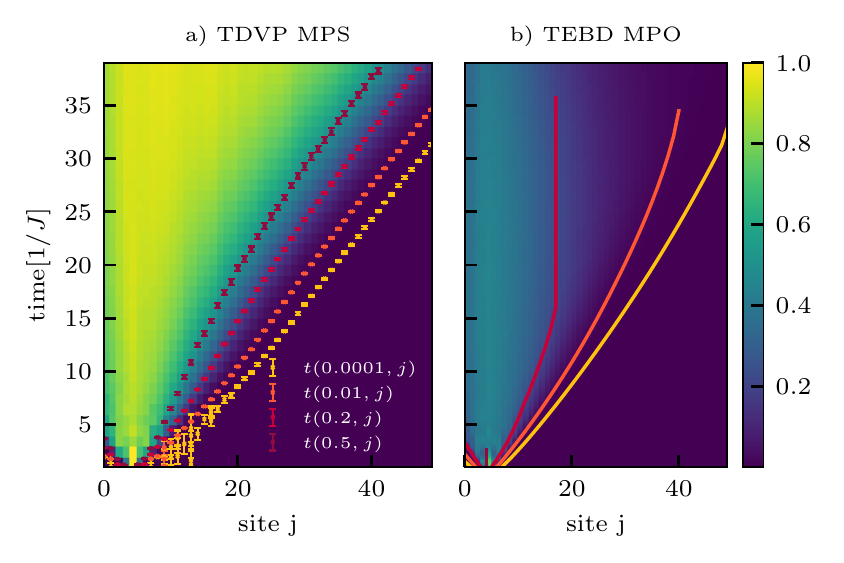}
\caption{OTOC $C_{4,j}(t)=\frac{1}{2}\|\left[ \hat \sigma_4^z(t), \sigma_j^z\right]\|_F^2$ as a function of distance and time for MPS TDVP with time step $\mathrm{d}t=0.005$, bond dimension $\chi=64$ and averaged over $387$ random states a) and MPO TEBD with time step $\mathrm{d} t=0.01$ and bond dimension $\chi=128$ b), both for system size $L=50$. The full lines and symbols correspond to contour lines obtained from the numerical solution of the equation $C_{4,j}(t) = \theta$ for various thresholds $\theta$, where $j$ is the position of the constant operator in the chain. The contour lines given by these solutions are denoted $t(\theta,j)$. The errorbars are representing the error coming from the stochastic sampling and are extracted using the bootstrap method.} 
\label{fig:light_cones} 
\end{figure}

\begin{figure}
\centering
\includegraphics{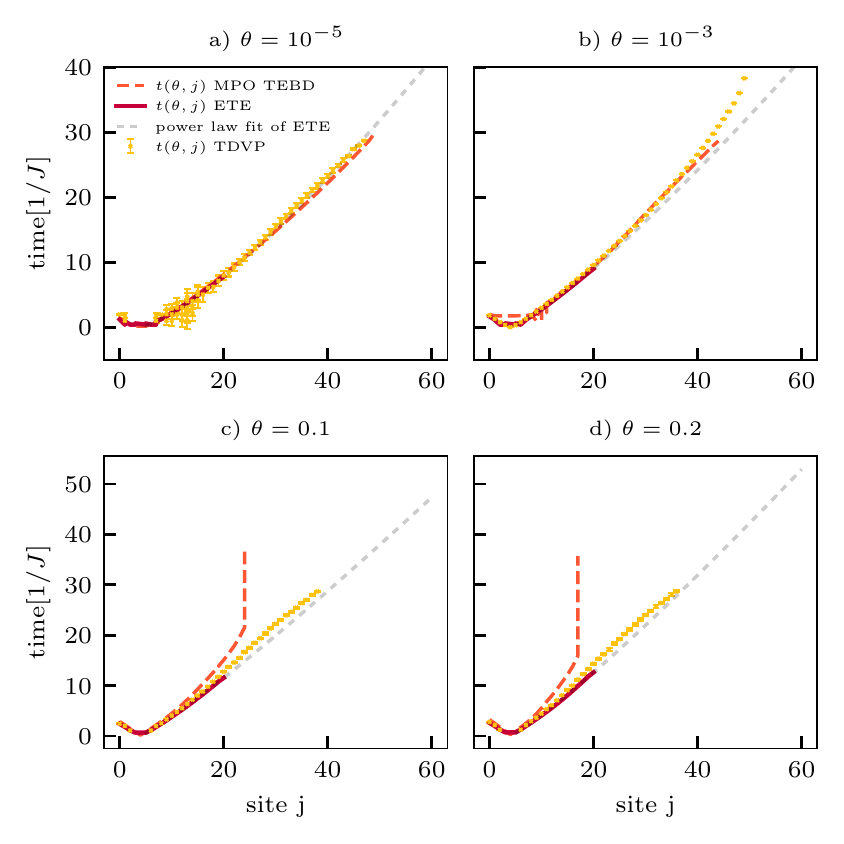}
\caption{Contour lines obtained from the numerical solution of the equation $C_{4,j}(t) = \theta$ for various thresholds $\theta$ and methods, where $j$ is the site. These solutions are denoted $t(\theta,j)$. The different methods used are MPO TEBD, MPS TDVP and ETE. We fit the ETE data between the sites $4$ and $21$ with a power law: $f(x)=\lambda(4-x)^\mu$ with $x$ the distance and the fitting parameters: a)$\lambda=0.20$, $\mu=1.32$, b)$\lambda=0.31$, $\mu=1.21$, c)$\lambda=0.50$, $\mu=1.12$, d)$\lambda=0.52$, $\mu=1.14$. The data comes from the same calculation as Fig. \ref{fig:light_cones} (MPS TDVP: bond dimension $\chi=64$, time step $\mathrm{d}t=0.005$, system size $L=50$ sites, averaged over 387 random states; MPO TEBD:bond dimension $\chi=128$, time step $\mathrm{d}t=0.01$, system size $L=50$ sites). The errorbars are obtained in the same way than in Fig. \ref{fig:light_cones} a.} 
\label{fig:comparison_contours}
\end{figure}

So far, we have presented results for systems small enough such that we could still compare to numerically exact results obtained by ETE. 
In what follows, we investigate the performance of these MPS and MPO methods for larger systems.
We present in Fig. \ref{fig:light_cones} the results for the OTOC $C_{4,j}(t)=\|\left[ \hat \sigma_4^z(t), \sigma_j^z\right]\|_F^2$ as a function of time $t$ and distance $j-4$ for a system of size $L=50$ sites with bond dimension $\chi=64$ for both MPO TEBD and MPS TDVP. Again, we chose the position of the spreading operator $\hat \sigma_4^z(t)$ on the left of our chain with open boundaries instead of on the center, since this allows for a better resolution of the tails of the right part of the OTOC as discussed in Ref. \cite{luitz_emergent_2018}.
For both methods, the time step $\mathrm{d}t$ has been decreased until convergence, and we found that a significantly smaller time step for TDVP of $\mathrm{d}t=0.005$ was required compared to the MPO TEBD time step of $\mathrm{d}t=0.01$, since the splitting methods of TDVP and TEBD differ. While in TEBD the exponential of the Hamiltonian is decomposed into two-site gates using a Trotter decomposition, in TDVP the update of every tensor requires the integration of coupled differential which are solved separately for every site $i$. In TEBD, only neighboring terms do not commute, while in TDVP the differential equations involving tensors $A^{[i]s_i}_{kl}(t)$ at site $i$ depend on the value of the tensors $A^{[j]s_j}_{kl}$ on all sites $j$. Therefore, a larger time step error can be expected in TDVP due to a more severe approximation in the splitting method.

For this reason, the dependence of the error on the time step is more important and must be checked carefully. Additionally, the stochastic sampling of the trace in Eq. \eqref{eq:otocdef} does not admit importance sampling and is therefore costly, practically limiting the bond dimensions considered here to $\chi=64$. 
We evaluate the convergence in bond dimension of our results for both methods in appendix \ref{sec:appendix_bonddim}.
We find that we achieve convergence for low values of the OTOC, which is consistent with the benchmarks shown in Fig. \ref{fig:panel_cuts}.

Here, we do not consider MPS TEBD results because of the inferior accuracy of this method already identified for smaller systems as discussed in the previous section. 

The representation of the results in Fig. \ref{fig:light_cones}  from the two methods on the same colorscale illustrates the problem observed for smaller systems in Fig. \ref{fig:panel_cuts} that MPO TEBD (right panel) underestimates the saturation value of the OTOC, in agreement with recent results of Ref. \cite{Hubig_review}, while TDVP (left panel) reproduces the correct long time saturation value close to $1$. Next, we consider contour lines (solid lines in Fig. \ref{fig:light_cones}) $t(\theta,j)$ of the OTOC obtained from numerical solutions of the equation $C_{4,j}(t) = \theta$ for various thresholds. For very low thresholds, these contours capture the behavior of the tail of the OTOC, where both methods yield consistent results even at long times. At larger thresholds, the obtained contours are strikingly different: while MPS TDVP yields approximately linear contour lines, close to a linear light cone, the results obtained with MPO TEBD deviate strongly and yield a significantly slower information spreading.
Due to the problems identified for MPO TEBD closer to the saturation regime of the OTOC as discussed in \ref{sec:benchmarks}, we attribute this behavior to the error caused by the insufficient amount of operator entanglement included in our MPO approximation.
The approximately ballistically spreading information front obtained with our MPS TDVP approach appears to be a qualitative improvement in comparison with MPO time evolution where the speed of information propagation seems to be underestimated. However, although qualitatively interesting, these results should not be trusted quantitatively at high thresholds since they are not converged in bond dimension in this region of space-time. 

The results displayed in Fig. \ref{fig:light_cones} can only be compared qualitatively, therefore we proceed by extracting the contours $t(\theta, j)$ of the OTOC for various values of the threshold and plot the results from MPS TDVP and MPO TEBD in the same figure panel for a direct quantitative comparison. In addition to the MPS results for $L=50$ also the exact results for $L=21$ are shown
in Fig. \ref{fig:comparison_contours}. This is an important comparison, since results in other systems demonstrate that for short enough times, the OTOC does essentially not show any finite size effects \cite{luitz_information_2017, luitz_emergent_2018}.
The contours obtained with ETE and MPS/MPO methods for different system sizes are therefore expected to coincide for short times and the contours at low thresholds should not depend on system size.

For very small thresholds ($\theta=10^{-5}$ and $\theta=10^{-3}$), see Figs. \ref{fig:comparison_contours}a and \ref{fig:comparison_contours}b, the contours obtained with MPO TEBD and MPS TDVP indeed match the exact results, in accordance with results of Fig. \ref{fig:panel_cuts}, confirming our expectations. However, some differences start to appear at higher thresholds ($\theta=0.1$ and $\theta=0.2$), see Figs. \ref{fig:comparison_contours}a and \ref{fig:comparison_contours}b, which can be expected from our study of convergence in bond dimension (see appendix \ref{sec:appendix_bonddim}). We note that our MPS TDVP seems to yield a contour slightly closer to the exact result.
 
For small thresholds, it was previously observed in generic spin systems that the contours of the OTOC assume a power law shape with exponents close to unity \cite{luitz_information_2017}. Therefore, we attempt power law fits to our numerically exact contours from ETE, yielding excellent fits. The fits are shown as gray dashed lines in Fig. \ref{fig:comparison_contours}, and should be understood as an extrapolation of the shape of the light cone from the $L=21$ results.
For small thresholds ($\theta=10^{-5}$), the MPS/MPO approaches reproduce the extrapolated contours with very high accuracy, confirming the power law fit from the smaller system size and consistency with the exact result. For $\theta=10^{-3}$, the two approximated approaches are still in quite good agreement with the fit of the ETE, although some differences arise at later times. At higher thresholds, the difference is even more significant, since already short times results do not agree. This confirms our overall observation that the MPS approaches considered here reproduce the tail of the OTOC with good accuracy, while the growth and saturation regimes are not well captured.
\section{Conclusion}

We have compared different MPS approaches to study information scrambling in a generic spin chain based on both matrix-product states (MPS) and matrix-product operators (MPO) and compared the results to an unbiased and numerically exact technique (ETE). For the calculation of the out-of-time-order correlators (OTOCs) in the Schr\"odinger picture based on MPS, we have shown that the use of a unitary time evolution method (TDVP) yields a significant improvement over the non-unitary truncation used in  the time evolving block decimation (TEBD) algorithm.
Furthermore we found that both MPO TEBD and MPS TDVP reproduce the tail of the OTOCs even at long times for low enough thresholds, while the growth and saturation regime suffers from truncation errors. The obtained shape of the light cone in large systems at low thresholds is in quantitative agreement with the ETE results at short times for smaller system size. Moreover they also match the extrapolated exact result even at late time.
For larger thresholds, closer to the information front, a discrepancy from exact results is observed, which we attribute to insufficient convergence of both MPO TEBD and MPS TDVP results with bond dimension. However, our TDVP MPS approach still yields a qualitatively correct ballistic propagation of information in contrast with the results obtained with MPO TEBD where after significant truncation the spreading of information appears to halt, making the result unphysical. We also note that the asymptotic saturation value of the OTOC is correctly reproduced in our TDVP MPS approach, while strong truncation effects in the MPO TEBD approach lead to a severe underestimation of the saturation value.

We conclude that both MPS techniques in the Heisenberg and Schr\"odinger picture yield consistent results for the tails of the OTOC and their performance is comparable. However, the MPS TDVP approach comes at the price of introducing a stochastic sampling of the OTOC using random product states making it computationally much more expensive.
An interesting future direction would be to apply our wave function approach to calculate the OTOCs in many body localized system to evaluate whether the logarithmic growth of entanglement allow us to gather reliable results in a broader region of space-time and calculate the contours OTOCs at larger thresholds.
\begin{acknowledgments}
	We would like the thank Michael Knap, Tibor Rakovszky, Johannes Hauschild and Yevgeny Bar Lev for the stimulating discussions.
	This project has received funding from the European Union’s Horizon 2020 research and innovation program under the Marie Sk\l{}odowska-Curie grant agreement No. 747914 (QMBDyn). We acknowledge PRACE for awarding access to
	HLRS’s Hazel Hen computer based in Stuttgart, Germany under grant number 2016153659. FP acknowledges the support of the DFG Research Unit FOR 1807 through grants no. PO 1370/2- 1, TRR80, the Nanosystems Initiative Munich (NIM) by the German Excellence Initiative, the DFG under Germany’s Excellence Strategy– EXC-2111-390814868, and the European Research Council (ERC) under the European Union’s Horizon 2020 research and innovation program (grant agreement no. 771537).
\end{acknowledgments}

\appendix

\section{Convergence with bond dimension}
\label{sec:appendix_bonddim}

\subsection{Convergence with bond dimension and comparison to exact results for TDVP MPS and TEBD MPS}
\begin{figure}
	\includegraphics{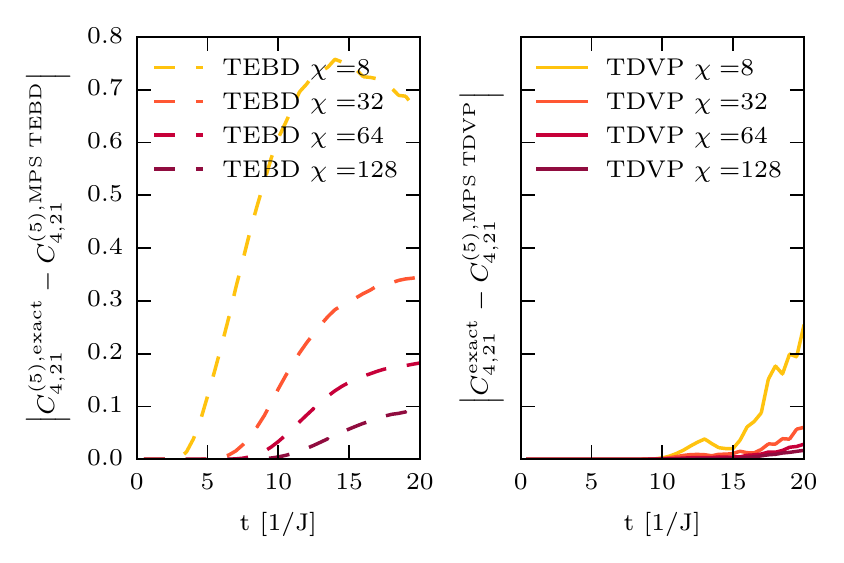}
    \caption{Comparison of the error $|C_{4,j}^{\mathrm{exact}}(t)-C_{4,j}^{\mathrm{TDVP} \; \mathrm{MPS}}|$ for TDVP MPS and TEBD MPS $|C_{4,j}^{\mathrm{exact}}(t)-C_{4,j}^{\mathrm{TEBD} \; \mathrm{MPS}}|$  for different bond dimension $\chi$. The system sizes is $L=21$, the time step is $\mathrm{d}t=0.01$.
	\label{fig:convergence L21}	
}
\end{figure}

In any MPS calculation, the convergence of the results with the bond dimension is important. Here, we present an analysis of the convergence of the OTOC results calculated by our stochastic TDVP and TEBD methods based on MPS. We also analyze the convergence with the bond dimension of our TEBD MPO results. 

In Fig. \ref{fig:convergence L21}, for a system of size $L=21$, we show the convergence of the OTOC $C_{4,j}(t)$ with bond dimension for the MPS time evolution methods that we compare in the main text (TDVP and TEBD). 
Since both methods rely on a stochastic sampling of the trace in Eq. \eqref{eq:otocdef}, we eliminate the error induced by the stochastic sampling by selecting 5 random product states $\ket{\psi_k}$ and then calculating the approximate OTOC $C_{4,21}^{(5)} = 1 - \frac{1}{2 \cdot 5} \sum_{k=1}^{5} \bra{\psi_k} \hat \sigma_4^z(t) \hat\sigma_j^z \hat\sigma_4^z(t) \hat\sigma_j^z \ket{\psi_k}$ with MPS TDVP, MPS TEBD and ETE for different bond dimensions (always using the same 5 product states). We plot the error given by the difference to the exact result for these $5$ states ($|C_{4,21}^{(5), \mathrm{exact}}-C_{4,21}^{(5), \mathrm{MPS}}|$) for different bond dimensions between $\chi=16$ to $\chi=128$ in Fig. \ref{fig:convergence L21}. For clarity, we indicate the number of random product states included in this comparison in parentheses (here by $(5)$). We observe a clear convergence of the results from MPS TDVP. For MPS TEBD the error also decreases with the bond dimension, but stays always much larger than the one of TDVP. This underlines the advantage of the conservation of unitarity by TDVP time evolution.
\begin{figure}
	\includegraphics{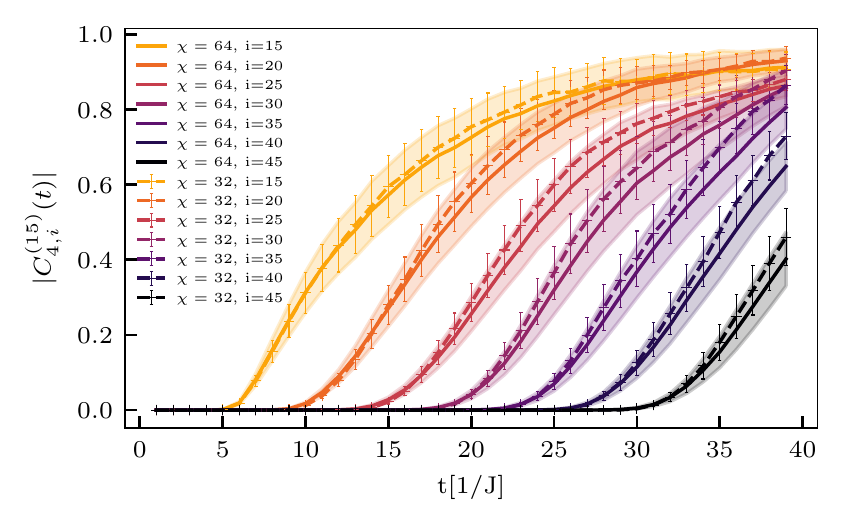}
	\caption{Convergence in bond-dimension of fixed space cuts of the OTOC $C_{4,j}^{(15)} (t) =1 - \frac{1}{2 \cdot 15} \sum_{k=1}^{15} \bra{\psi_k} \hat \sigma_4^z(t) \sigma_j^z \sigma_4^z(t) \sigma_j^z \ket{\psi_k}$ obtained with Schr\"odinger TDVP with the same 15 randomly chosen initial product states $\ket{\psi_k}$ for bond dimension $\chi=64$ and $\chi=32$ for time step $\mathrm{d}t=0.01$, and system size $L=50$.}
	\label{fig:convergence contours}
\end{figure}
\subsection{Convergence with bond dimension for larger systems in TDVP MPS}

\begin{figure}
	\includegraphics{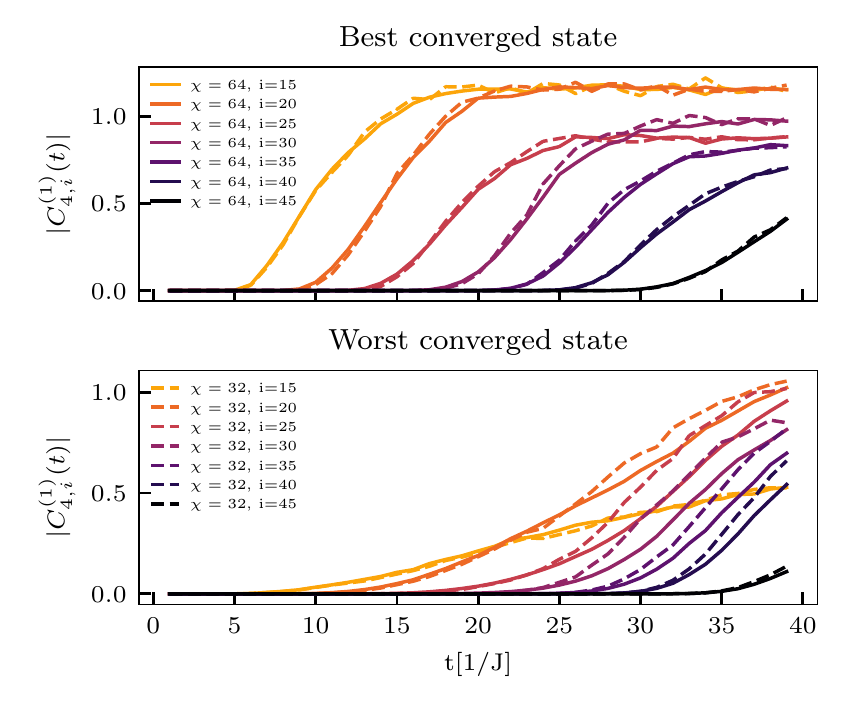}
    \caption{Convergence in bond-dimension of fixed space cuts of the OTOC $C_{4,j}^{(1)}(t)$ obtained with Schr\"odinger TDVP for only \emph{one} random initial product state for bond dimension $\chi=64$ and $\chi=32$ and using a time step $\mathrm{d}t=0.01$ for a system of size $L=50$. The worst and best converged initial states of the ones used in Fig. \ref{fig:convergence contours} are displayed in order to demonstrate the difference in convergence depending on the initial state.}
	\label{fig:best_worst}
\end{figure}

For larger system sizes, it is difficult to obtain exact results for the OTOC as a benchmark.
Therefore, a careful analysis of the dependence of the results on the bond dimension is crucial to identify the domain of validity of the methods. 
\begin{figure}
	\centering
	\includegraphics{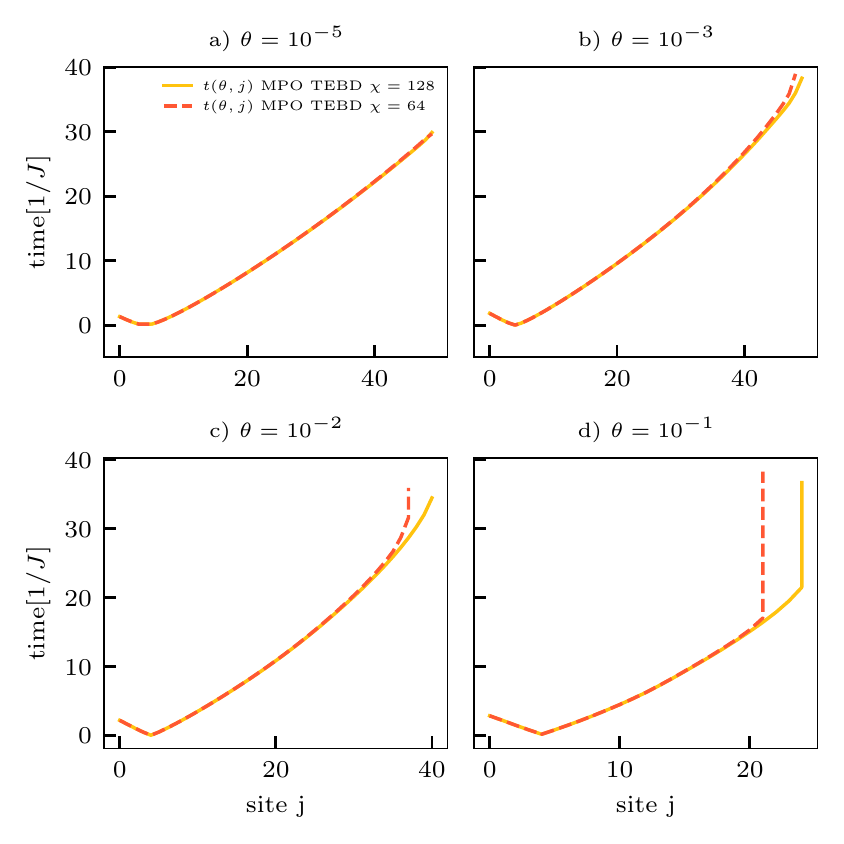}
	\caption{Convergence with bond dimension $\chi$ of the contours lines of the OTOCs calculated with MPO TEBD. We represent the numerical solution of the equation $C_{4,j}(t) = \theta$ for various thresholds $\theta$ and bond dimension ($\chi=64$ and $\chi=128$) for system size $L=50$ sites and time step $\mathrm{d}t=0.01$. $j$ is the site of the light cone for different thresholds $\theta$ (see Fig. \ref{fig:comparison_contours})}
	\label{fig:convergence chi MPO}
\end{figure}

In the case of the TDVP MPS approach, we compare cuts $C_{4,j}^{(15)}(t)$ of the approximate OTOC using 15 random initial product states in the $\sigma_z$ basis for several fixed distances. Note the same initial states are chosen for both bond dimensions in order to eliminate the importance of statistical errors in this comparison as explained above. For converged results within this approach, the mean and the error (calculated using bootstrap sampling) obtained for different bond dimension results should perfectly agree. 
In Fig. \ref{fig:convergence contours}, we show the approximate OTOC $C_{4,j}^{(15)}(t)$ for $\chi=32$ and $\chi=64$ together with the errorbars of the OTOC (shaded region for $\chi=64$ and errorbars for $\chi=32$), yielding very good agreement of the results for low thresholds. At larger thresholds, the discrepancy between the two bond dimension results becomes significant as expected.

We find that the convergence in bond dimension depends significantly on the initial state and therefore we repeat this analysis for approximate OTOCs $C_{4,j}^{(1)}(t)$ using only single random product states and different bond dimensions in Fig.\ref{fig:best_worst}. From the 15 product states included in Fig. \ref{fig:convergence contours}, we select the states with the best and worst convergence in bond dimensions to illustrate these state to state differences. Overall convergence is only achieved only for low values of the OTOCs, confirming the observation that the tail of the OTOC is reproduced accurately, while values at larger thresholds are not converged.

\subsection{Convergence with bond dimension for larger systems in TEBD MPO}

In the case of the MPO TEBD approach, the study of the convergence in bond dimension is facilitated by the absence of stochastic sampling. The spacio-temporal dependency of the effect of bond dimension can be analyzed by directly looking at extracted contour lines $t(\theta,j)$ of the OTOC obtained from numerical solutions of the equation $C_{4,j}(t) = \theta$ for various thresholds and bond dimensions. We present the result of this approach in Fig. \ref{fig:convergence chi MPO}.
The contours obtained with different bond dimension coincide very well for small thresholds (up to $\theta=10^{-3}$). However a difference between $\chi=64$ and $\chi=128$ appears already for $\theta=10^{-2}$ at times $t=30$, which we attribute to the insufficient representation of the operator entanglement of the operator in an MPO with $\chi=64$. The difference is even more striking for $\theta=0.1$, where the results are not converged and do not show the expected asymptotic behavior of a linear light cone. The breakdown appears earlier for the lowest bond-dimension as expected.

\bibliography{otoc_tdvp}
\end{document}